\documentstyle[preprint,revtex]{aps}
\begin{document}
\draft
\preprint{UFIFT-HEP-92-10\\
          hep-ph/9210210}
\begin{title}
Wiggly Relativistic Strings
\end{title}
\author{Jooyoo Hong$^*$, Jaewan Kim and Pierre Sikivie }
\begin{instit}
        Department of Physics,
        University of Florida,
        Gainesville, FL 32611
\end{instit}
\begin{abstract}
We derive the equations of motion for general strings, i.e.
strings with arbitrary relation between tension $\tau$ and energy
per unit length $\epsilon$.  The renormalization of $\tau$ and
$\epsilon$ that results from averaging out small
scale wiggles on the string is obtained in the general case to
lowest order in the amount of wiggliness.
For Nambu-Goto strings we find deviations from the
equation of state $\epsilon \tau
= {\rm constant}$ in higher orders.
Finally we argue that
wiggliness may radically modify the gauge cosmic string scenario.
\end{abstract}
\pacs{PACS numbers:}
\narrowtext
Computer simulations \cite{simulation} have shown that cosmic
gauge strings acquire much more small scale structure than had been thought
previously \cite{vilenkin85}.
The effect of small scale structure is generally speaking to
increase the energy per unit length of a gauge string and to decrease
its tension when a coarse grained description of the wiggly
string is adopted \cite{carter90,vilenkin90}.
While this fact has been appreciated in the
past, it appears that there is no general treatment yet of the
renormalization of string tension and energy per unit length due
to small scale wiggliness.  It appears even that there is no
relativistic description in place for the motion of strings
when the tension differs from the energy per unit length.
The present paper aims to fill this gap.  The formalism
developed below can be applied to all strings and we hope it will
prove useful in other contexts.  An example which comes to mind
is a gauge string with fermion zero modes attached to it \cite{witten84}.
When a large number of particles are attached to a
string, they act collectively as a fluid adding to the energy
per unit length but subtracting from the tension.

Consider then a general string, i.e.\ an object whose
stress-energy-momentum is localized on a line in space.  Let
$X^\mu (\sigma)$ be the location of the string worldsheet with
respect to a Lorentz reference frame.  $\sigma = (\sigma^0,\sigma^1)$
are arbitrarily chosen coordinates parametrizing
points on the worldsheet.  The associated 2-dim. metric is, as
usual:
\begin{equation}
h_{ab} (\sigma) =\partial_a X^\mu \partial_b X_\mu\ ,~~~a,b=0,1~.
\label{metric}
\end{equation}
At point $X^\mu (\sigma)$ on the string lives a 2-dim.
stress-energy-momentum tensor
\begin{equation}
t^{ab} = (\epsilon - \tau) u^a u^b + \tau~h^{ab}, \label{emtensor}
\end{equation}
where $\epsilon (\sigma)$ is the energy per unit length of the
string, $\tau (\sigma)$ its tension and $u^a (\sigma)$ is the
fluid velocity parallel to the string; $u^a u_a = +1$.  The 4
dim. stress-energy-momentum tensor is
\widetext
\begin{equation}
T^{\mu\nu} (x) = \int d^2\sigma \sqrt{-h}~
t^{ab} (\sigma) \partial_a X^\mu (\sigma) \partial_b X^\nu (\sigma) \delta^4
\Bigl( x-X (\sigma)\Bigr)\ , \label{4emtensor}
\end{equation}
\narrowtext
\noindent where $h = \det(h_{ab})$.
This expression is both invariant under 2-dim. reparametrizations
and covariant under 4-dim.  Lorentz transformations.  Its
validity in the neighborhood of each point on the worldsheet can
be verified explicitly by choosing a Lorentz frame which is
instantaneously at rest with respect to the string at that point.
The motion of the string must be such that
$\partial_\mu T^{\mu\nu}(x) = 0 $.
It is easy to show that this condition is equivalent to
\begin{equation}
\partial_a \Biggl[ \sqrt{-h}~ t^{ab} (\sigma)
\partial_b X^\mu \Biggr] = 0~~~(\mu=0,1,2,3)\ . \label{4eqmo}
\end{equation}
To provide a complete description of the string dynamics, Eq.(\ref{4eqmo})
must be supplemented by an equation of state:
\begin{equation}
\tau = \tau (\epsilon)\ . \label{eqst}
\end{equation}
We then have five equations for the five unknowns
$\vec X_\perp (\sigma),~\epsilon (\sigma),~\tau (\sigma)$ and $\beta (\sigma)
\equiv u^1 (\sigma)/u^0 (\sigma)$.  $\vec X_\perp (\sigma)$
represents the two transverse degrees of freedom of the string.
$\beta (\sigma)$ is its longitudinal velocity.  Eqs.(\ref{4eqmo}) and
(\ref{eqst}) uniquely specify the motion of the string that results from
arbitrary initial conditions.  Note that our description is
Lorentz invariant as well as generally covariant in the 2-dim.
sense.

The case of the Nambu-Goto (NG) string is included in this
description.  The well-known NG equations of motion are Eq.(\ref{4eqmo})
with $t^{ab} (\sigma)$ replaced by $h^{ab} (\sigma)$.  Let us
first note that two of these equations are merely mathematical
identities since
\begin{equation}
\partial_e X_\mu \partial_a \Biggl[ \sqrt{-h}~ h^{ab}
(\sigma) \partial_b X^\mu \Biggr] = 0~~~(e=0,1) \label{iden}
\end{equation}
follows from Eq.(\ref{metric}) and nothing else.  The other two NG
equations specify the motion of $\vec X_\perp (\sigma)$.  Now,
returning to our description of a general string, let us adopt
the equation of state $\tau = \epsilon$.  It is easy to show,
using Eqs.(\ref{iden}), that two of the Eqs.(\ref{4eqmo}) are equivalent
to
\begin{equation}
\partial_a \tau (\sigma) = 0~~~~~~a = 0,1\ . \label{triv}
\end{equation}
They therefore imply that $\tau$ must be a constant.  The two
remaining Eqs.(4) are the non-trivial NG equations that
determine the motion of $\vec X_\perp (\sigma)$.  Thus we have
learned that $\tau={\rm constant}$ is the only consistent way to have
$\tau = \epsilon$.

The equation of state appropriate to a particular kind of string
must be derived from the relevant microphysics.  As in other
studies of fluid dynamics, an average over a large number of
micro-configurations consistent with a given macroscopic
description must be performed to obtain the energy per unit
length $\epsilon$ and the tension $\tau$ of the string.  The
equation of state gives the relationship between $\epsilon$ and
$\tau$ when the string is slowly stretched (i.e. the stretching
timescale is long compared to the microphysics timescales).  The
focus of our paper is the renormalization of $\epsilon$ and
$\tau$ due to small scale wiggles on a string with arbitrary
equation of state.  We must define an averaging length scale
$\lambda \equiv {2\pi\over k}$.  $\epsilon (k)$ and $\tau (k)$
are the values of the energy per unit length and tension of the
string when all wiggles of wavelength shorter than $\lambda$
are averaged over.  We will derive the
renormalization group equations for $\epsilon$ and $\tau$ due to
transverse and longitudinal wiggles to
second order in the amplitude of the wiggles.

Consider a string which lies on average along the $x$-axis.  We
choose $t$ and $x$ as worldsheet coordinates.
Thus $ \Bigl(X^\mu(\sigma)\Bigr) = \Bigl( t,x,y(t,x),z(t,x)\Bigr)$.
In this gauge, Eqs.(\ref{4eqmo}) are:
\begin{mathletters}
\begin{equation}
\partial_a \biggl( \sqrt{-h} \,t^{ab} \biggr) = 0\, ,
\end{equation}
\begin{equation}
t^{ab} \partial_a \partial_b y = t^{ab} \partial_a \partial_b z= 0\ .
\end{equation}
\end{mathletters}
\noindent Let us first discuss transverse wiggles.  Consider a string of
equation of state $\tau = \tau (\epsilon)$ stretched along the
$x$-axis.  At rest $(\beta=y=z=0)$ the string has energy per unit
length $\epsilon_{_0}$ and tension $\tau_{_0} = \tau
(\epsilon_{_0})$.  At
time $t=0$, the string is given a transverse velocity in the
y-direction:  $\dot y(0,x) = \beta_{_T}\sin kx$.  We assume
$\beta_{_T} \ll 1$ and expand Eqs.(8) in powers
of $\beta_{_T}$. This yields to lowest order
\begin{mathletters}
\begin{eqnarray}
\epsilon_{_0} \ddot y - \tau_{_0} y^{\prime\prime} &=& 0 \label{timev1}\\
\ddot \epsilon_{_1} - v_{_L}^2 \epsilon_{_1}^{\prime\prime} &=&
(-\epsilon_{_0} \partial_t^2 + \tau_{_0}\partial_x^2) {1\over
2} (\dot y^2 + y^{\prime 2}) \label{timev2} \\
(\epsilon_{_0} - \tau_{_0}) \dot\beta &=& \tau_{_0} y^\prime (\ddot y -
y^{\prime\prime}) + v_{_L}^2 \epsilon_{_1}^\prime \label{timev3}
\end{eqnarray}
\end{mathletters}
\noindent for the time evolution of $y$, $\beta$ and
$\epsilon_{_1} = \epsilon -\epsilon_{_0}$.  By definition
$v_{_L}^2 \equiv -{d\tau\over d\epsilon}\mid_{_0}$.
As usual, dots and primes denote derivatives
with respect to $t$ and $x$.  Eq.(\ref{timev1}) implies
$y =(\beta_{_T}/w)\sin kx \sin\omega t$ with $\omega = k\,v_{_T}$
where
$v_{_T} =({\tau_0\over \epsilon_0})^{1/2}$ is the phase velocity of
transverse wiggles.  Eqs.(\ref{timev2}) and (\ref{timev3}) determine
$\epsilon_{_1} (\sigma)$ and $\beta (\sigma)$, both of which are of
order $\dot y^2$.  To obtain the renormalized values
$\bar\epsilon_{_T}$ and $\bar \tau_{_T}$ of the energy per unit length
and tension, we calculate $\langle T_{\mu\nu} (x) \rangle$ to
second order in $\dot y$.  The result, including the contribution
from wiggles in the $x$-$z$ plane, is

\begin{mathletters}
\begin{eqnarray}
\bar\epsilon_{_T}&=&\epsilon_{_0}+\epsilon_{_0}
(\langle \dot y^2\rangle +\langle \dot z^2\rangle ) \label{twiggle1} \\
\bar\tau_{_T}&=&\tau_{_0}-{1\over 2}(\langle \dot y^2 \rangle +
\langle \dot z^2 \rangle )\Biggl[\tau_{_0} + \epsilon_{_0} + v_{_L}^2
\epsilon_{_0}
\biggl( 1-{\epsilon_{_0}\over \tau_{_0}} \biggr)\Biggr]\,.\nonumber\\
&&\label{twiggle2}
\end{eqnarray}
\end{mathletters}
\noindent Next, let us discuss longitudinal wiggles.
Again we start with a string which is at rest, stretched along the
$x$-axis.  In this state, it has energy per unit length
$\epsilon_{_0}$ and tension $\tau_{_0} = \tau (\epsilon_{_0})$.  At time
$t=0$, the string is given a longitudinal velocity $\beta (0,x) =
\beta_{_L}\sin kx$.  Provided \hbox{$\beta_{_L} \ll 1$},
Eqs.(9) are still
valid but now $y=z=0$.  They imply that
\hbox{$\epsilon_{_1}\sim(\epsilon_{_0}-\tau_{_0})v_{_L}^{-1}\beta_{_L}$}
and that
\hbox{$v_{_L} = ( -{d\tau\over d\epsilon}\mid_{_0} )^{1/2}$} is the
phase velocity of longitudinal wiggles.  For the renormalization
of $\epsilon$ and $\tau$ due to longitudinal wiggles we find, up
to second order in $\beta$:

\begin{mathletters}
\begin{eqnarray}
\bar \epsilon_{_L} &=& \epsilon_{_0} + 2(\epsilon_{_0} - \tau_{_0}) \langle
\beta^2\rangle \label{lwiggle1} \\
\bar\tau_{_L} &=& \tau_{_0} - (\epsilon_{_0} -\tau_{_0})
\langle\beta^2\rangle \Biggl[ 1+ v_{_L}^2 + (\epsilon_{_0} -
\tau_{_0})
{d\ln v_{_L}\over d\epsilon}\bigm|_{_0}\Biggr]\ .\nonumber  \\
&&\label{lwiggle2}
\end{eqnarray}
\end{mathletters}
\noindent Thus the promised renormalization group equations for
$\epsilon(k)$ and $\tau (k)$ are

\begin{mathletters}
\begin{eqnarray}
- {d\epsilon\over d \ln k} & = & W_{_T} (k) \epsilon + 2 W_{_L} (k)
(\epsilon - \tau) \\
- {d\tau\over d \ln k} & = &-{1\over 2} W_{_T} (k)
\Biggl[\tau + \epsilon + v_{_L}^2 \epsilon (1- {\epsilon\over
\tau})\Biggr]    \\
&&\qquad -W_{_L}(k)(\epsilon -\tau)\Biggl[1+v_{_L}^2+(\epsilon -\tau)
{d \ln v_{_L}\over d\epsilon} \Biggr].  \nonumber
\end{eqnarray}
\end{mathletters}
\noindent where $W_{_T} (k)$ and $W_{_L} (k)$ are the spectral densities on a
$\ln k$ scale of respectively $\langle \dot y^2\rangle +
\langle \dot z^2\rangle$ and $\langle \beta^2\rangle$.
Eqs.(12) relate the values of $\epsilon$ and $\tau$
at one scale to their values at a vastly different scale provided that
$W_{_L},\,W_{_T}\ll 1$ at all intermediate scales.  Note that we have not yet
obtained how the equation of state itself changes from scale to
scale.  To do so we need to analyze the response of the wiggles
to adiabatic stretching of the string.  We leave this to a future
publication which will also contain the details of the
calculation that led to Eqs.(10-12).

Let us consider the case of wiggly Nambu-Goto strings.  At the
shortest distance scale $k_0$ we have $\epsilon = \tau =\mu$,
where $\mu$ is the bare string tension.  At slightly longer
distance scales Eqs.(12) imply $\epsilon (k) = \mu [1+W_{_T}
(k_{_0})\ln k_{_0}/k]$
and $\tau (k) = \mu [1-W_{_T} (k_{_0}) \ln k_{_0}/k]$.
Therefore in the neighborhood of $k=k_0$ we have the equation of
state $\epsilon\tau = \mu^2$.  Moreover, a short calculation
shows that Eqs.(12) imply ${d(\epsilon\tau)\over d\ln k} = 0$
when the equation of state is $\epsilon\tau =$ constant.  This
equation of state is therefore a fixed point of the
renormalization group equations~(12).  Thus in the particular
case of wiggly Nambu-Goto strings, Eqs.(12) do by themselves
establish the equation of state to be $\epsilon\tau = \mu^2$ in
lowest order.  (As was already emphasized, for a general string
the renormalization group equations for $\epsilon$ and $\tau$ do
not by themselves provide enough information to determine the
equation of state.  The Nambu-Goto string is an exception in this
regard.)  The equation of state $\epsilon\tau =\mu^2$
was found earlier by Carter \cite{carter90} and Vilenkin \cite{vilenkin90}.
It seems that we have found considerable support for it.  However we will
now show that in general $\epsilon\tau \not= \mu^2$ for wiggly
Nambu-Goto strings although $\epsilon\tau = \mu^2$ may be an
excellent approximation in many cases.  The fact that
$\epsilon\tau \not= \mu^2$ in general does not contradict what we
have said so far because Eqs.(12) are valid only in lowest
order.

It is well known that an arbitrary motion of a Nambu-Goto
string \cite{GGRT} is given by
${\bf x} (t,\sigma) = {1\over 2} [{\bf a} (\sigma -t) +
{\bf b} (\sigma + t)]$ where ${\bf a}$ and ${\bf b}$ are arbitrary
functions subject only to the constraint ${\bf a}^{\prime 2} =
{\bf b}^{\prime 2} = 1$.  To describe a wiggly NG string lying on
average along the $x$-axis we write
\widetext
\begin{eqnarray}
{\bf a} (\sigma -t) &= \Bigl[ (\sigma - t) \gamma_{_1} + f_{_1}
(\sigma - t),~g_{_{1y}} (\sigma - t),~g_{_{1z}} (\sigma - t)\Bigr]\nonumber  \\
{\bf b} (\sigma + t) &= \Bigl[ (\sigma + t) \gamma_{_2} + f_{_2} (\sigma +
t),~g_{_{2y}} (\sigma + t),~g_{_{2z}} (\sigma +
t)\Bigr]  \label{ngstring}
\end{eqnarray}
\narrowtext
\noindent where $\gamma_{_1}$ and $\gamma_{_2}$ are constants and
$f_{_1},~f_{_2},~g_{_{1y}},~g_{_{1z}},~g_{_{2y}}$ and $g_{_{2z}}$ are
functions which average to zero and which describe the wiggles on
the string.  These functions are not all independent since they
must obey the gauge condition
${\bf a}^{\prime 2} = {\bf b}^{\prime 2} = 1$.  Let us
choose ($t,x$) as the worldsheet coordinates of the averaged
string.  It is easy to show that in these coordinates the 2-dim.
stress-energy-momentum tensor of the averaged string is given by
\widetext
\begin{eqnarray}
t^{00} &=& \mu \biggl\langle {1\over \mid{dx\over
d\sigma}\mid}\biggr\rangle = \mu \biggl\langle{2\over
\gamma_{_1}+\gamma_{_2}+f_{_1}^\prime
(\sigma - t)+ f_2^\prime (\sigma + t)}\biggr\rangle \nonumber \\
t^{01} &=& \mu\biggl\langle{-\gamma_{_1}+\gamma_{_2} - f_{_1}^\prime
(\sigma - t) + f_2^\prime
(\sigma + t)\over \gamma_{_1} + \gamma_{_2} + f_{_1}^\prime (\sigma - t) +
f_{_2}^\prime (\sigma + t)} \biggr\rangle    \\
t^{11} &=& -2\mu \biggl\langle {(\gamma_{_1}+f_{_1}^\prime (\sigma
-t))(\gamma_{_2}+f_{_2}^\prime (\sigma +t))\over
\gamma_{_1}+\gamma_{_2}+f_{_1}^\prime (\sigma
-t)+f_{_2}^\prime (\sigma +t)}\biggr\rangle  \label{ngemtensor} \nonumber
\end{eqnarray}
\narrowtext
\noindent Clearly $\epsilon\tau = -t^{00} t^{11} + (t^{01})^2 \not= \mu^2$
in general.  For example, consider the particular case where the
wiggles are reflection-symmetric on average $(\gamma_{_1} =
\gamma_{_2} \equiv \gamma,~\langle f_{_1}^{\prime p}\rangle = \langle
f_{_2}^{\prime p} \rangle \equiv \langle f^{\prime p}\rangle$ for $p =
2,3,\ldots)$.  Then an expansion of $t^{ab}$ in powers of $f^\prime$
yields:
\widetext
\begin{eqnarray}
   \epsilon  & = t^{00} & = {\mu\over \gamma} \Biggl[ 1+ {1\over 2\gamma^2}
 \langle f^{\prime 2}\rangle - {1\over 4\gamma^3}\langle f^{\prime 3}\rangle
 + {1\over 8\gamma^4} (\langle f^{\prime 4}\rangle + 3\langle
 f^{\prime 2}\rangle^2)-\cdots\Biggr]   \nonumber \\
   \tau      & = -t^{11} &= \gamma\mu \Biggl[ 1 - {1\over 2\gamma^2}\langle
 f^{\prime 2}\rangle + {1\over 4\gamma^3}\langle f^{\prime 3}\rangle
 -{1\over 8\gamma^4}(\langle f^{\prime 4}\rangle - \langle f^{\prime 2}
 \rangle^2) + \cdots\Biggr].   \label{parameter}
\end{eqnarray}
\narrowtext
\noindent Eqs.(\ref{parameter})
show that $\epsilon \tau \not= \mu^2$ although
the $2^d$ and $3^d$ order terms in the
expansion of $\epsilon\tau - \mu^2$
vanish.  That the $2^d$ order term
vanishes, we already knew from Eqs.(12).  In general,
one has
\begin{equation}
{\epsilon\tau\over\mu^2} =
1+{1\over 4\gamma^2}\langle f_{_1}^{\prime 2}\rangle
\langle f_{_2}^{\prime 2}\rangle +\cdots .
\end{equation}
Note that if the wiggles are purely transverse
$(f_{_1}^\prime =f_{_2}^\prime =0)$ then $\epsilon \tau = \mu^2$ to all
orders \cite{vilenkin90}. However, it is
clear that one must allow $f_{_1}^\prime,~ f_{_2}^\prime \not=0$.
Physically this corresponds to the possibility of longitudinal wiggles
once the
Nambu-Goto string has $\epsilon > \tau$ because of
transverse wiggles.

Finally, we would like to speculate on the behavior of cosmic
gauge strings.  Let us assume that higher order terms in the
renormalization group Eqs.(12) do not play an important role.
The equation of state is then $\epsilon \tau = \mu^2$, and
\begin{equation}
- {d\epsilon\over d\ln k} = W_{_T}\epsilon + 2W_{_L} (\epsilon -
{\mu^2\over \epsilon})\ .\label{cosmic_rge}
\end{equation}
How large are $W_{_T}$ and $W_{_L}$?  At cosmic time t, when
the correlation length of the string
network is $\xi (t)$, the strings carry wiggles which have been
inherited from earlier times when the correlation length was
shorter.  For $\lambda = {2\pi\over k}$ somewhat
shorter than $\xi (t)$, wiggles are abundant
and the corresponding values of $W_{_T}$ and
$W_{_L}$ are large, of order one.  For $\lambda < G\mu t$,
$W_{_T}$ and $W_{_L}$ are exponentially suppressed because the
decay time of wiggles on the bare string into gravitational
radiation is of order $(G\mu)^{-1} \lambda$ \cite{gravrad}.
For $G\mu t <\lambda < \xi (t)$,
the values of $W_{_T}$ and $W_{_L}$ are the
outcome of a number of competing processes \cite{processes}
some of which tend to
increase and some of which tend to decrease the size of wiggles
associated with the corresponding length scales.    Stretching of the
strings (e.g. by Hubble expansion) and the production of loops by
self-intersection with reconnection tend to decrease the size of
wiggles, whereas shortening of the string after reconnections
have occurred and the production of kinks, also as a result of
reconnections, tend to increase the size of wiggles.
We will assume here that $W_{_T}$ and $W_{_L}$ have approximately
constant values for all $\lambda$:  $G\mu t < \lambda < \xi (t)$.
We make this assumption not because we believe that it is necessarily
correct but as a means to explore the effect of small scale wiggliness
on the cosmic gauge string scenario. With regard to the computer
simulations, it is unclear to us whether they are in disagreement with
this assumption. There is so far no published result describing
unambiguously the spectrum of wiggliness.
%
%

At any rate, if $W_{_T}$ and $W_{_L}$ are
approximately constant for $G\mu t <\lambda < \xi (t)$, then from
Eq.(\ref{cosmic_rge}),
\begin{equation}
\epsilon = \epsilon (\xi)= {\mu^2\over \tau} \sim \mu
\biggl({\xi (t)\over G\mu t}\biggr)^{W_T+2W_L}.
\end{equation}
The typical velocity of cosmic strings is then
\begin{equation}
v(t) =
\biggl[{\tau (t)\over \epsilon (t)}\biggr]^{1\over 2} \sim
\biggl({G\mu t\over \xi (t)}\biggr)^{W_T+2W_L}\, .
\end{equation}
One expects the correlation length to be $\xi (t)=v(t)t$.  This
determines $\xi (t) \sim t(G\mu)^\alpha$ where $\alpha =
{W_T+2W_L\over W_T+2W_L+1}$.  The density of strings today is
then given by:
\begin{equation}
\Omega_{\rm str} \sim (6\pi Gt^2){\epsilon \over \xi^2 (t)} \sim
(6\pi)(G\mu)^{1-3\alpha}\ .\label{density}
\end{equation}
Because of limits on the anisotropy of the microwave background
radiation, $\Omega_{\rm str}$ must be much smaller than one.  This
requires $\alpha < {1\over 3}$ or $W_{_T}+2W_{_L} < 0.5$.
Even if this condition is satisfied, the limit on $G\mu$ may be
much more severe than it is in the usual scenario which assumes
$\alpha =0$.  For example, if $W_{_T}+2W_{_L} = 0.2,~\Omega_{str}
< 10^{-5}$ implies $G\mu < 10^{-12}$ instead of $G\mu < 10^{-6}$
in the usual scenario.

\acknowledgments
We are grateful to A. Vilenkin, P. Ramond, C. Thorn, T. Kibble,
E. Copeland and P. Griffin for useful discussions.  One of us
(P.S.) would like to thank the Aspen Center for Physics for its
hospitality while he was working on part of this project.
This work was supported in part by the U.S.\ Department of Energy
under contract DE-FG05-86ER40272, and by the Korea Science and
Engineering foundation.

\end{document}